# Physical principles of building protein megacomplexes in a crowded milieu

Authors: Jiayi Wang[1], Jules Nde[2], Andrei G. Gasic[3], Jacob Haseley[1], Margaret S. Cheung[1,4*]
Affiliations:
1. Department of Physics, University of Washington, Seattle, WA, USA
2. Department of Cancer Biology, University of Kansas Medical Center, Kansas City, KS, USA
3. Geminus.ai, Inc, Houston, TX, USA
4. Environmental Molecular Sciences Laboratory, Pacific Northwest National Laboratory, Richland, WA, USA

- Corresponding author: margaret.cheung@pnnl.gov

**Abstract**
Multiple phenotypic protein expressions arise from one genome represent variations in the protein relative abundance and their stoichiometry. A lack of definite compositional parts challenges the modeling of protein megacomplexes and cellular architectures. Despite the advance in protein structural predictions with AI, the mechanism of protein interactions and the emergence of megacomplexes they assemble remains unclear. Here, we present a statistical-physics framework of grand canonical ensemble to explore the protein interactions that drive the emergent assembly of a megacomplex using the observational mass spectrometry datasets including protein relative abundance and the cross-linked connections. Using chromatin remodeler megacomplex, INO80, as an example, we discovered a class of "divergent" protein that plays a critical role in orchestrating the assembly beyond nearest neighbors, dependent on the excluded volumes exerted by others. With the constraints of the excluded volumes by varying crowding contents, these divergent subunits orchestrate and form clusters with selective components growing into configurationally distinct architectures. We propose a machinery view for the INO80 chromatin remodeler complex where each loosely associated subunits can be occasionally recruited for parts as attachment into a core assembly driven by excluded volumes. Our computational framework provides a mechanistic insight into taking the macromolecular crowding as necessary physicochemical variables representing cell states to remodel the configurations of protein megacomplexes with structurally loose modules.

**Significance statement:**

Protein complexes dynamically remodel into megacomplexes in a crowded cell. However, the physical principles governing the configurational organization of protein parts into megacomplexes in response to a cell state remain unclear. Here, we developed a statistical physics framework to gain insights into the spatial organization of proteins shaped by excluded volumes, inter-protein interactions, and relative abundance. By characterizing the contribution of individual subunits toward assembling a megacomplex, we highlight the importance of incorporating cell-state variables, such as crowding and relative protein abundance, in determining the configuration of a megacomplex with multiple loosely attached parts.

# I. Introduction

There has been a rapidly growing interest in understanding why many cellular phenotypes arise from one genome in the content of environmental gradients enabled by high-throughput experimentations, imaging, and advance analytic tools guided by AI [1]. Given that genomic information and derived knowledge is abundantly available, we gained insight about biological networks from gene expression data [2-4], single-cell RNA sequencing [5], and protein sequences and structures [6]. However, the multiple phenotypic expressions at the proteomic level involving variations in protein interactions and their abundance representing a cell state [7] remain poorly understood (**Figure 1A**) [8, 9]. Protein assemblies that dynamically adapt to a cell state [10, 11] by associating into biomolecular condensates [12-16] or megacomplexes[17-20] are outstanding examples that manifest a growing complexity from genotype to phenotype expressions shaped by environmental influences.
Despite the advance in protein structural predictions with AI, the mechanism of protein interactions and the emergence of megacomplexes they assemble remains unclear.

To address this gap, several computational and theoretical frameworks have been developed [12-16, 21-23] to investigate the protein interactions that drive macromolecular assemblies involving heterotypical constituents. One approach is the scaffold-client framework, in which proteins interact with more partners (i.e. higher valency) are considered scaffolds to recruit the clients interacting with few partners (i.e. lower valency) [24, 25]. While there are extended models to include more comprehensive interactions between scaffolds and clients, this framework is still under the assumption of a fixed protein level [26, 27], representing a static cell state.

To incorporate the consideration of dynamic cell states that influence the constituents as well as the configurations of protein assemblies, we deployed an open system based on a grand canonical ensemble framework [28] to test a hypothesis that the excluded volumes of subunits limit the configurational search of protein interactions in a reservoir-like cytoplasm environment, and that drives protein assemblies. We used yeast chromatin remodeler INO80 complex [29-35] as an example for the assembly of a megacomplex with many structurally loose parts in a cell-like environment and they remodels [29, 30] upon environmental cues.

We represented a dynamic cell state where a system continuously exchanges components and moves toward equilibrium with the surrounding particle reservoir. We generated structural models by inferring interaction energies and chemical potentials between loose parts from experimental data for INO80. Through thermodynamic perturbations such as abundance variation and crowding [36-39], we teased out the critical components in INO80 which orchestrates interactions beyond nearest neighbors. The excluded volumes contributed to the depleted interactions [40] between modules at high crowding content, giving rise to distinct configurational state of structural machinery for INO80 under various crowding conditions (**Figure 1B**).

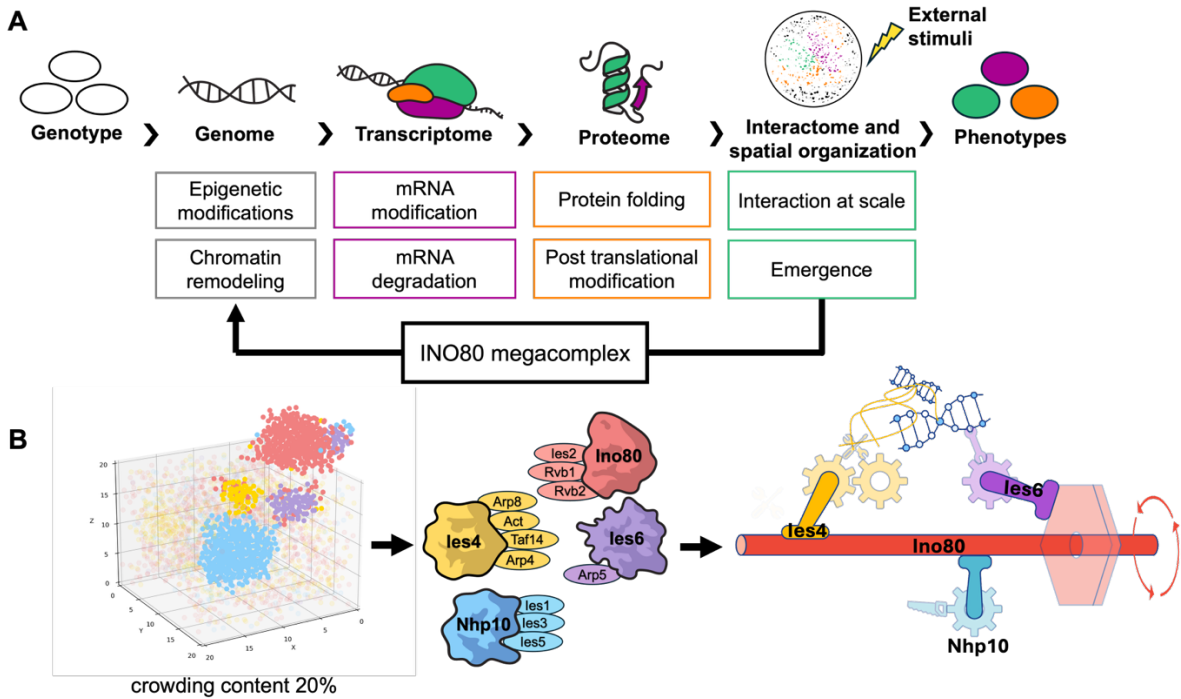

**Figure 1**. Protein spatial organization mechanisms as a knowledge gap. (A)Transition from genotype to phenotype, where understanding of proteins' spatial arrangement under external stimuli is critically lacking. The INO80 chromatin remodeling complex, which modifies chromosome topology according to cell state in the yeast metabolic cycle, is chosen as a candidate to gain insight into the protein's spatial organization. (B)The simulated cluster formation at a specific volume fraction and the inferred subunits' hierarchical order in spatial regulation. Red, purple, blue, and yellow represent subunits in the Ino80 core, Arp5, Nhp10, and Arp8 modules, respectively. Cloud and round shapes represent the difference in hierarchy during subunit gathering.

## II. Method

### A. Composition of INO80 complex.

Yeast INO80 consists of 15 subunits(see supplement for more details), which are grouped into four modules (bold represents modules): the **Ino80 Core** (Ino80, Ies2, Rvb1, Rvb2), **Arp8** (Arp8, Arp4, Actin, Ies4, Taf14), **Arp5** (Arp5, Ies6), and **Nhp10** (Nhp10, Ies1, Ies3, Ies5) modules; each module has its distinct function[31-33, 41-44].

### B. Inversely determine INO80 subunits' interaction energies and chemical potentials

To model the spatial-temporal process of INO80 complex cluster formation in a grand canonical ensemble (**Figure 1B**), we coarse-grain each protein as a bead of the same size. The following input variables are needed: the relative abundance of subunits, interaction energies between subunit types, and chemical potentials. The relative abundance of INO80 subunits can be measured experimentally. However, the interaction energies and chemical potentials can only be inferred from empirical values using a maximum entropy approach, in which we estimate subunit interaction energies and chemical potentials from protein abundance and interaction data.

For the relative abundance of INO80 subunits, we take data from the Mass Spectrometry study in Sardiu 2015[45]. The total number of particles in the simulation box is then determined using the equation $N_{total} = L^3\phi/v_0$, where $L$ is the length of the cubic box, $\phi$ is the volume fraction, and $v_0 = \frac{1}{6}\pi\sigma^3$ is the volume of a single bead with $\sigma$ being the diameter of the beads. We use a box size of $L=20\,\sigma$. The total particle number is then distributed into multiple particle types according to their relative abundance.

For inter-protein interactions, we infer the interaction energies that could reproduce the empirical observations from the cross-linking experiment. The Lennard-Jones potential $U_{ijab}(r) = \frac{\epsilon_0}{|r_i^a - r_j^b|^{12}} - \frac{\epsilon_{ij}}{|r_i^a - r_j^b|^6}$ is used to model the interaction between two beads of type $i$ and $j$; $a$ and $b$ are the indices of particle, ranging from 1 to $N_i^{exp}$, and 1 to $N_j^{exp}$ correspondingly. $r$ is the coordinate vector representing the location of each particle. The repulsive interaction energy $\epsilon_0$ is universal to all beads to account for excluded volume, while the attractive interaction energy $\epsilon_{ij}$ depends on bead types.

We first estimate an inter-protein contact intensity for each pair of subunit types from the cross-linking experiment in Tosi 2013 [34]. Then, using the algorithm developed by Zhang and Wolynes 2015[46], we iteratively update the interaction energies $\epsilon_{ij}$ until the simulated inter-protein contact intensities match the experiment observations to a certain extent. To identify the most representative simulation trajectory of INO80, we also consider the known INO80 modules and screen over the range of simulation trajectories that meet the previous selection criteria. The detailed procedure, mathematical derivation, and the resulting interaction energies are shown in the supplement.

After the interaction energies are determined, we proceed to determine each subunit's chemical potential. To do that, we build on the maximum entropy algorithm developed in Gasic 2021[47], and enable the subunit's particle exchange with the reservoir one at a time. For every single subunit, we iteratively update the chemical potential while fixing the abundance for the other 14 subunits, until the particle number of the investigated subunit in the simulation matches the experimental observation:

$$\mu^{k+1} = \mu^k - k_l[\langle N \rangle - \widetilde{N}]^k$$

(1)

where μ is the chemical potential of the subunit of inquiry, $k$ is the iteration number, and $k_l$ is an adjustable learning rate. $\langle N \rangle$ is the ensemble average of the subunit's simulated particle number at the current iteration, and $\widetilde{N}$ is the empirical abundance. The percentage difference $\frac{\langle N \rangle - \widetilde{N}}{\widetilde{N}}$ is calculated after each iteration; the iteration is stopped when the percentage difference is below 0.01.

### C. Simulation of subunits assembly and analysis

Depending on whether a convergence to empirical abundance can be reached, subunits are classified into two types: convergent and divergent (see the results section). To simulate the formation of the INO80 megacomplex, convergent subunits freely conduct particle exchange with the cellular reservoir, while the abundances of divergent subunits are held strictly constant. The INO80 assembly is thus modeled in a Grand Canonical Monte Carlo simulation where a fixed number of divergent subunits recruit convergent subunits from the reservoir based on their chemical potentials. We then investigate the physics principles underlying the convergent and divergent properties to elucidate their role in subcellular spatial organization through the following steps:

*1. Mapping chemical potential's relation with particle number*

To understand the fundamental physics behind convergence and divergence behavior, we performed simulations over a range of chemical potentials for a single subunit type, while keeping the particle numbers of the others fixed at the empirical values. We then map the subunit's particle number against chemical potentials. Among the fifteen subunits, we focused on those exhibiting divergent behavior in at least one volume fraction and analyzed their profiles across all three volume fractions.

To probe the physical basis of this convergence-divergence phenomenon, we also perturbed the temperature to T = 0.9 and T = 1.1 at a fixed volume fraction of 0.05 and examined the resulting abundance–chemical potential curves. For volume fraction 0.05, we use 1.5 million Monte Carlo steps. For volume fractions 0.1 and 0.2, we use 2 million Monte Carlo steps and multiple initial conditions when needed.

*2. Particle number perturbation*

To better understand how each subunit's particle number shapes the assemblage, we performed canonical Monte Carlo simulations (CMC) using empirical particle numbers for all 15 subunits. Then, each subunit's count was varied by ±10, 20, 30, or 40, together with a test particle with only volume exclusion and no interaction with others for baseline reference. For low-abundant subunits Ies4 and Ies6, perturbations exceeding their empirical counts were marked "N/A." All simulations were run at reduced temperature $T = 1$ and volume fraction $\phi = 0.1$ for 4,000,000 timesteps, dumping 4,000 frames (1 frame per 1,000 steps). Analyses used every 10th frame from 2,000–4,000 to sample equilibrated trajectories. We analyze the shifts in the ensemble averages of network features (see Supplement) and thermodynamic properties for each perturbation.

For thermodynamic properties, we chose two indicators – change in accessible volume $\Delta V_{acc}$[48, 49] and potential energy $\Delta E$. Accessible volume is defined as the total space available for a test particle of a given size to be inserted. Consider inserting a test particle into a system of $i$ particles, we have:

$$V_{excluded} = \bigcup_i \left[\frac{4}{3}\pi(r_i + r_{test})^3\right]$$
$$V_{acc} = V_{tot} - V_{excluded}$$

Where $V_{excluded}$ is the excluded volume, and $V_{acc}$ is the accessible volume. $r_i$ and $r_{test}$ are the radius of the existing and test particles, and $V_{tot}$ is the total volume of the system (in this case, the volume of the box).

*3. Calculating the effective potential mean force (PMF)*

To further distinguish the system-wide effects of divergent and convergent subunits, we use Ovito to compute the radial distribution function $g(r)$ between selected subunits and all particles in the simulation box. We then derived the corresponding effective potential mean force (PMF) using the relation $W(r) = -K_B T \ln g(r)$.

## III. Results
### A. A small subset of subunits shows "divergent" characteristics in reservoir-like cytoplasm

We model the cellular impact of relative protein abundance into chemical potential[47] and their interactions from cross-linking mass spectrometry for the INO80 complex (see Method section). For each fifteen subunits of the INO80 complex, we iteratively adjust the chemical potential until the simulated particle number converges to its experimental abundance while keeping the other fourteen subunits' abundance fixed. For most subunits, their convergence is reached within a few iterations (see Supplement Figure S4). However, for a small subset of subunits, the particle number oscillates around the experimental abundance and are unable to converge toward experimental estimates, regardless of hyperparameters. We call this specific category of subunits "divergent". Others are by comparison "convergent".

To better understand the underlying principles of the newly discovered *divergent* property of subunits in INO80, we plotted the phase diagram of density vs. chemical potential of the subunits that display divergent behavior at a reduced temperature unit, T=1, and the volume fractions of subunits, $\phi$ equals 0.05, 0.1, and 0.2 in **Figure 2**. This range of $\phi$ represents the volume fractions of crowding contents in a cell[38, 50]. Density represents normalized abundance and $\phi$ is defined as the proportion of a system's total volume that is occupied by the INO80 subunits. In **Figure 2A**, the profiles for the divergent subunits are plotted in thick lines. They exhibit a discontinuity in density around a critical chemical potential $\mu^*$. For example, the profile of Ies4 (yellow curve) at $\phi = 0.1$ shows a density discontinuity spanning from roughly 0 to 0.014 at chemical potential $\mu^* = -9.8$, within which the empirical value (indicated by the yellow arrow) falls. Similarly, Ino80 also exhibits this discontinuity in density that encompasses its empirical value (indicated by the red arrow). Both Ies4 and Ino80 display consistent divergence property across all volume fractions.

Interestingly, not all subunits display divergent behavior across the whole range of $\phi$. Some subunits, such as Ies6, Arp5 and Nhp10, exhibit emergent divergent property only within a narrow range of volume fraction. For example, Ies6 subunit is convergent at $\phi = 0.05$ but become divergent at $\phi = 0.1$ and $\phi = 0.2$. Arp5 and Nhp10, on the other hand, show critical behavior between divergent and convergent (more in Discussion section). Arp5 only become divergent at $\phi = 0.1$, and Nhp10 does so at $\phi = 0.2$. We highlight them with dotted lines in Figure 2A for visual guidance.

The discontinuity in the chemical potential vs density curves in **Figure 2A** of a divergent subunit indicates that the system is highly unstable within that region when connecting the subunit to a particle reservoir. A thermodynamic interpretation of the divergence phenomenon is provided using the framework of spinodal decomposition (**Inset of Figure 2A**), in which the particle numbers of divergent subunits fall within the unstable region of the phase transition (more in Discussion).

### B. Effective stickiness between divergent subunits and others emerges under crowding

Divergent subunits, with their empirical abundance (pointed by arrows in **Figure 2A**) being within this discontinuity region, are more suitably modeled with a fixed particle number rather than within a reservoir. We thus simulated the INO80 megacomplex formation using the empirical abundance of divergent subunits and the inferred chemical potentials of the rest across the three volume fractions (see Method). The formed clusters and their compositions are shown in Supplemental **Figure S6**. Interestingly, we observed that all formed clusters consist of at least one divergent subunit.

The divergent property of a subunit, however, depends on crowding, as manifested in Ies6. Given that subunit Ies6 subunit is convergent at $\phi = 0.05$ but become divergent at the volume fractions of 0.1 and 0.2, we further explore its interactions with other subunits at these varying content crowding by computing the effect potential of mean force (PMF). In **Figure 2B**, we derived PMF between Ies6 and all other subunits at the three volume fractions, $\phi$ of 0.05, 0.1 and 0.2, in order to understand the many-body interactions on the properties of Ies6 subunits (see Method section). At $\phi = 0.05$ there is only volume repulsion as PMF > 0 everywhere; at $\phi = 0.1$, an attractive well appears, since PMF < 0 when r, the distance between Ies6 and other subunits, is less than $5\sigma$ (the diameter of a subunit). Interestingly, when $\phi = 0.2$, PMF first decrease to below 0 at short r, and then increase above 0 when $6\sigma < r < 9\sigma$, representing a desolvation barrier separating the subunits. The presence of desolvation barrier adds the entropy cost to break apart interactions [51], essentially "strengthening" the depleted [40], sticky effect.

The changes in the profiles of effective PMF with $\phi$ demonstrates the fascinating phenomena where a subunit's effective interaction with the system is shaped by crowding content. Take Ies6 in Figure 2B as an example, it becomes divergent at $\phi \geq 0.1$, corresponding to the fact that it shows stickiness in the effective PMF as opposed to repulsion due to volume exclusion at $\phi = 0.05$. We observed that this behavior can be generalized to other subunits in **Figure S8**, which shows that effective PMF for the divergent subunits have an attractive potential well interacting with all other subunits at $\phi = 0.1$. In contrast, the convergent subunits exhibit only volume exclusion with all other subunits.

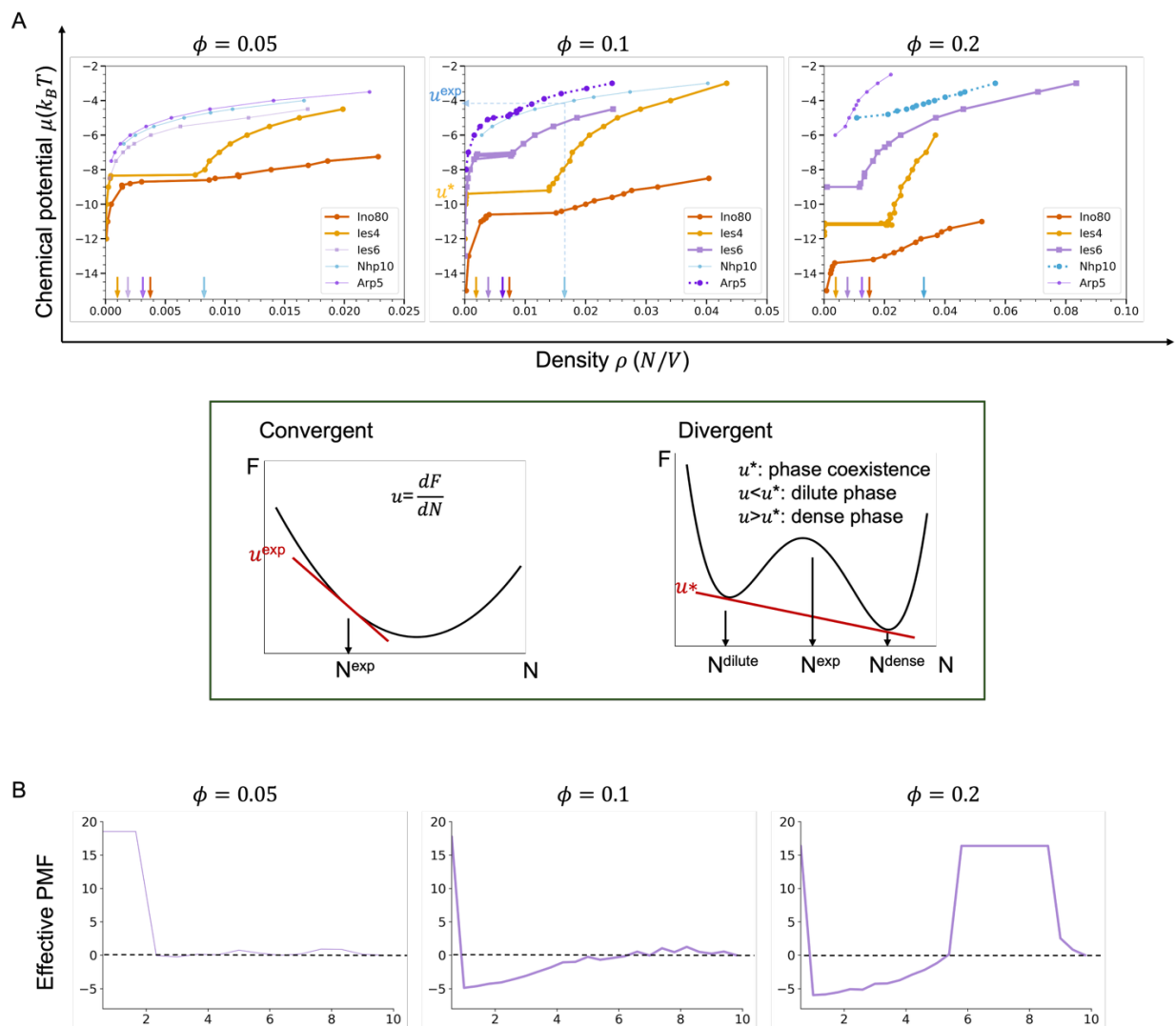

**Figure.2** Chemical Potential vs Density Profile and Potential Mean Force (PMF) for Subunits Showing Divergent Property at Any Volume Fraction. (A) Chemical potential vs. density for subunit Ino80, Ies4, Ies6, Nhp10, and Arp5 at volume fractions $\phi$ = 0.05, 0.1, and 0.2 and temperature T=1. Colors represent the INO80 module a subunit belongs to. Opaque curves indicate divergent subunits, and transparent curves indicate convergent subunits. Curves with dashed lines represent subunits at the boundary between convergence and divergence. The arrows on the x-axis represent each subunit's experimental density. The inset shows schematic diagram of the system's free energy landscape along the particle number of a single subunit with divergent or convergent property. The idea is adapted from the theory of spinodal decomposition[52]. (B) Effective potential mean force between Ies6 subunit and all particles at volume fractions $\phi$ = 0.05, 0.1, and 0.2 and temperature T=1.

### C. Divergent subunits orchestrate spatial organization beyond local interactions

To further understand the thermodynamic contribution of divergent subunits to the aggregation of assembled clusters, we analyzed the emergent behavior and thermodynamic properties by quantifying each subunit's contribution to the assembly they form. We calculated the free energy changes ($\Delta G$) represented by accessible volume ($\Delta V_{acc}$) and potential energy ($\Delta E$) resulting from variations in its particle number, $dN$ (see Method section). $V_{acc}$ is expressed in the unit of $\sigma^3$, and $E$ is in the unit of $k_B T$, where $k_B$ is the Boltzmann's constant.

To achieve a baseline understanding of particle insertion free energy, $\Delta G$, we measured $\Delta V_{acc}^0$ and $\Delta E^0$ using a test particle with only volume exclusion. We then compared it to the perturbation by inserting particles of each subunit $i$ from INO80 complex as $\Delta V_{acc}^i$ and $\Delta E^i$ by calculating the difference in perturbation $\Delta\Delta V_{acc}^i = \Delta V_{acc}^i - \Delta V_{acc}^0$ and $\Delta\Delta E^i = \Delta E^i - \Delta E^0$. The result is shown in **Figures 3.** From **Figure 3A**, we see that compared to test particles, inserting divergent subunits increases the accessible volume ($\Delta V_{acc}^i - \Delta V_{acc}^0 > 0$ at $dN > 0$) and removing them reduces accessible volume ($\Delta V_{acc}^i - \Delta V_{acc}^0 < 0$ at $dN < 0$). An insertion of a "sticky" divergent subunit induces nucleation and opens up the available volume for particle insertion by promoting effective attractive interactions with others, significantly lowering the potential energy as shown in **Figure 3B**. They facilitate cluster formation through long range interaction, as manifested in the long tail of an effective potential of mean force. In contrasts, perturbation of particle insertion from convergent subunit remains insignificant, and the impact of convergent subunits on others is simply local.

To theoretically describe the thermodynamic contributions of divergent and convergent subunits to the overall free energy change upon particle insertion, we break down into four terms (**Figure 3C**) in unit of $k_B T$. First, there is a direct influence from the conjugate of particle number and chemical potential ($\mu \Delta N$); second, there is a change in the potential energy ($\Delta E$); third, the conjugate of accessible volume $\Delta V_{acc}$, and the volume fraction $\phi$ ($-\phi \Delta V_{acc}$)[53]. Finally, the contribution in the configurational entropy from remodeling the spatial arrangements ($-\Delta S$) [36, 37]. We deduce that divergent subunits contribute more to the energy, entropy, and accessible volume terms, while convergent subunits contribute primarily through the change in particle number, represented by the $\mu \Delta N$ term.

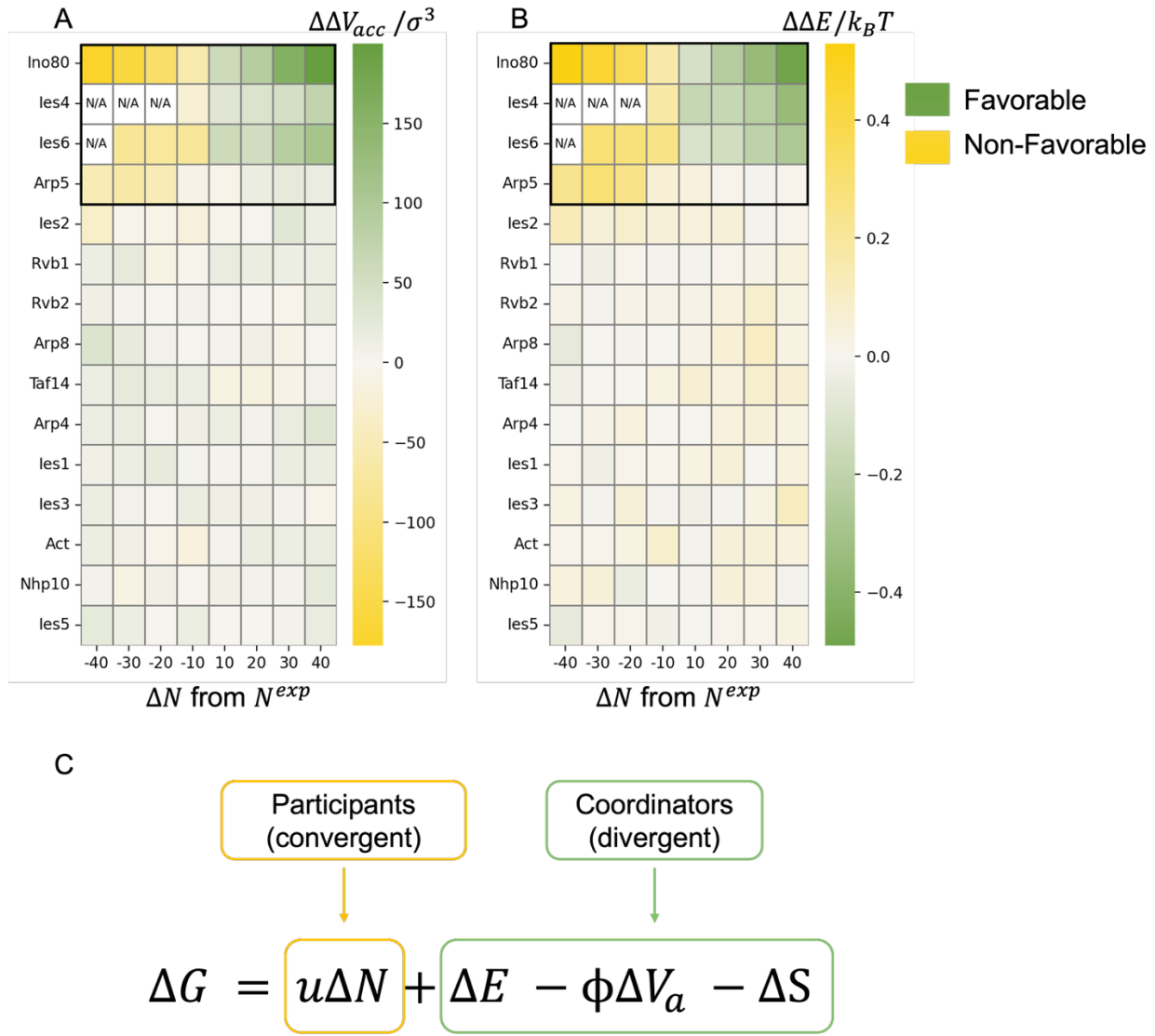

**Figure 3.** Free Energy Change from Particle Number Variation of Each INO80 Subunit. (A, B) Normalized change in accessible volume $\Delta\Delta V_{acc}^i$ and potential energy $\Delta\Delta E^i$ for each subunit $i$, respectively, caused by particle number variation at volume fraction $\phi = 0.1$ and temperature $T = 1$. Row represents subunit, and columns correspond to particle number perturbation $\Delta N$. Green indicates thermodynamically favorable responses, while yellow denotes thermodynamically unfavorable ones. Divergent subunits are in the black box. N/A represents an unphysical negative particle number due to low experimental abundance. (C) A theoretical illustration of the divergent and convergent subunits' contribution to the free energy terms.

### D. Interactions, abundance, and crowding contents corroborate the emergence of a divergent subunit

With the mechanistic understanding of subunit assembly, we propose to predict whether a subunit is divergent or not from only protein-protein interactions and the abundance -- both can be measured by proteome mass spectrometry experimentally. We provided a visual guidance of this prediction with a bidirectional, half-edged graph in **Figure 4A**. Because a pair of subunits, $i$ and $j$, may each have distinct protein abundance ($N_i^{exp} \neq N_j^{exp}$), we assigned the interactions as a pair of "source" and "target" into double-headed arrows showing an imbalanced impact to one another due to the many-body effect they each experience. The direction of an arrowhead points from a source subunit $i$ to a target subunit $j$, and the thickness of the half-edge represents interaction strength weighted by the abundance of the source: $\epsilon_{ij} N_i^{exp}$.

By observing this graph in **Figure 4A**, we identified the characteristics in the effective potential mean force as well as the uneven abundance between the pair that contribute to the directional impact on a target subunit, leading to its divergence at certain crowding content (see **Table S2**). For example, the number of heterotypic interacting partners is important to the emergence of divergent to Ino80 (bold in Figure 4A). It has the highest number of partners, and is divergent at $\phi = 0.05, 0.1$ and $0.2$. For another example, low abundance is a contributing factor for Ies4 (bold in **Figure 4A**). It has the lowest abundance, and is divergent at $\phi = 0.05, 0.1$, and $0.2$. For Ies6, its divergence is contributed by a combination of relatively strong interaction and relatively low abundance (bold in **Figure 4A**). Ies6 shows divergence at $\phi = 0.1$ and $0.2$.

In summary, we observed phenomenologically that the divergence of a target subunit satisfies two criteria: (1) when the interaction pointing towards a target subunit $j$ ($I_j = \epsilon_{ij} N_i^{exp}$) has outweighed its abundance ($N_j^{exp}$). (2) When the crowding exists and its $\phi$ reaches a certain threshold. We define a ratio of interaction to abundance $\frac{I_j}{N_j^{exp}}$ for each subunit $j$. The emergence of a divergent subunit is inferred by a high ratio value in **Table S2**. The higher the ratio, the lower the crowding content is required for a subunit to become divergent. For example, when $\frac{I_j}{N_j^{exp}} > 40$ (e.g. Ies6 and Ino80), then divergent happens at the crowding content as low as $\phi = 0.05$. When $\frac{I_j}{N_j^{exp}} > 30$ (Ies4) the threshold of crowding to be divergent increases to $\phi = 0.1$.

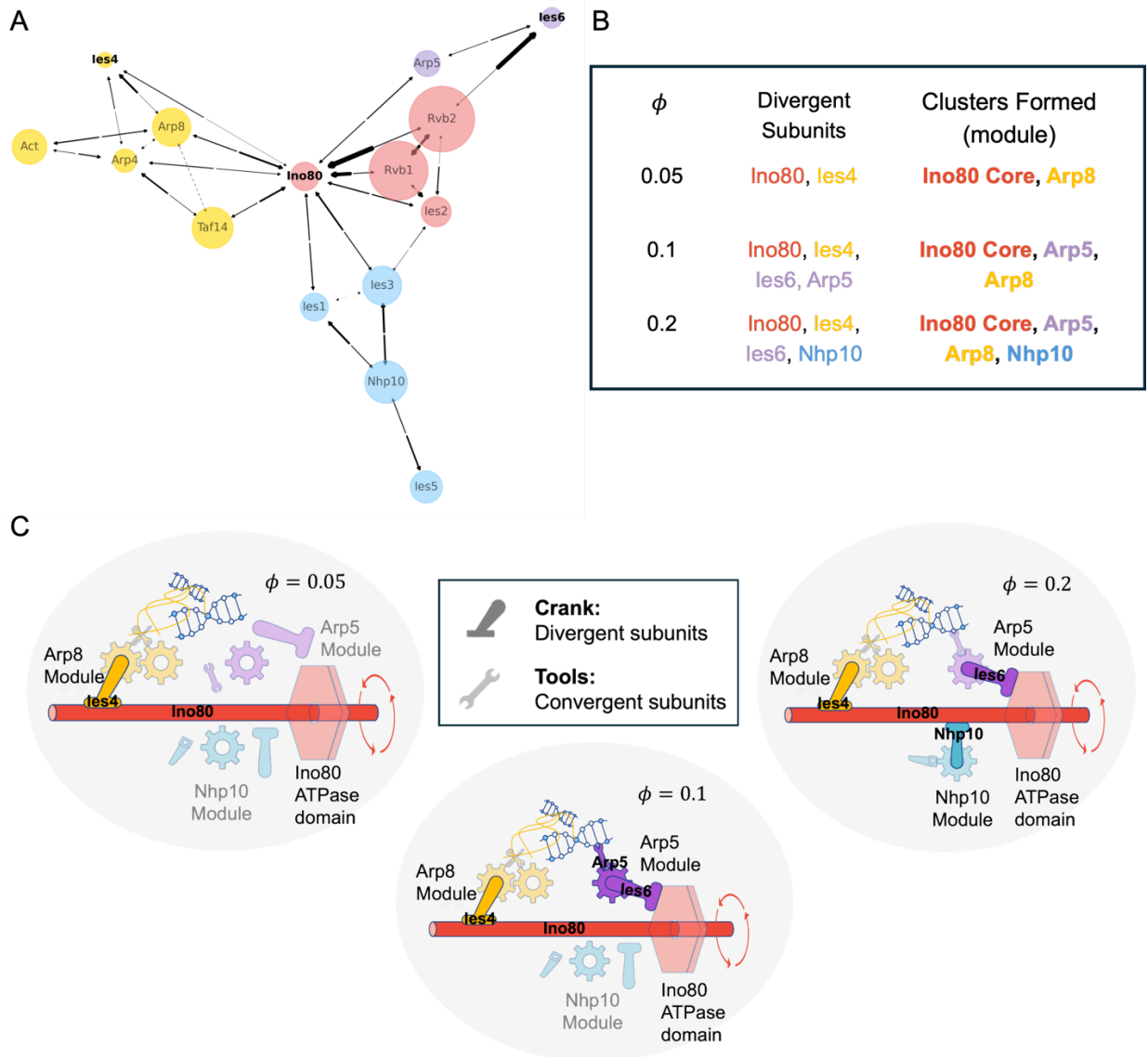

**Figure 4.**

Divergent Subunits Orchestrate Megacomplexes into Operational Machinery upon Crowding Changes
(A) A bidirectional, half-edge graphs of interaction networks weighted by subunit abundance for INO80 remodeling complex. Each node corresponds to a subunit, with half-edges indicating interactions with another subunit from the XL-MS experiment. Edges are directed to represent the weighted impact from one ("source) to the other ("target"); its width reflects the interaction strength ($\epsilon_{ij}$) weighted by the abundance of the source. The solid edges represent attractive interactions between the source and the target. The only repulsive interactions are represented by a dashed edge between Arp8 and Taf14. Node size reflects subunit abundance, and node color represents modules in the INO80. (B) Divergent subunits and cluster composition at various volume fractions summarized from Figure S6. Bold texts marks INO80 modules, shown in distinct colors; regular texts indicate subunits, colored according to the modules they belong. (C) A mechanical interpretation of the INO80 remodeling complex in the content of DNA and actin filaments at various volume fractions. Convergent subunits are tools that remodel DNA. Divergent subunits are viewed as cranks that operationally joins the tools together at the correct location along the

DNA and actin filaments to perform remodeling functions. The emergence of divergent properties for a subunit depends on crowding content ($\phi$ = 0.05, 0.1, 0.2). We represent the emergent divergent subunits in opaque, while the convergent ones are transparent.

**Discussion**
  A. **The emergence of a divergent subunit is contributed by system's heterogeneity and many-body effect**

The investigation of protein assemblies in crowded, cell-like environments remains challenging. One major difficulty is that the energy landscape governing the assembly of heterotypic subunits under crowding is highly frustrated [54], which obscures the underlying physical principles at the cellular level [CMS2]. To reduce this complexity, many existing models assume uniform interactions in a canonical ensemble. For example, Sanders et al. (2020) [27] modeled stress-granule and P-body composition by capping subunit binding sites under uniform interaction strengths and fixed protein abundance. While this framework successfully reproduced certain compositional features, it did not explore how variations in interaction heterogeneity or abundance reshape the phase behavior. For the same system, Jacob et al. [26] introduced a "bridge" particle to study wetting transitions as functions of interaction strength and stoichiometry, again within a canonical ensemble with uniform interactions. Although they solved the phase diagram under these assumptions, they did not examine alternative cellular states with different protein abundances, and thus their theory probed only a narrow region of phase space. As a result, it is difficult to assess how their conclusions extend to more realistic, compositionally diverse assemblies.

To determine what phase space is actually required to represent protein assemblies in vivo, we use realistic interaction and abundance data derived from XL-MS and proteomics. Building on the grand-canonical ensemble approach introduced by Gasic and co-workers[47], and by Sartori and Leibler[55], who extended phase space over interaction energies and chemical potentials (protein abundances) to show that crowding can drive protein assembly, we adopt a similar statistical-mechanical framework. We then apply this framework to a realistic chromatin assembly structure, a system known for its dynamic, heterotypic interactions in a crowded, cell-like environment. This allows us to probe frustration and organization in a physiologically relevant megacomplex, thereby bridging a critical gap between idealized theoretical models and experimentally derived structures.

Within this framework, we discovered a surprising "divergent" characteristic of certain subunits, in contrast to others that behave "convergently." This divergence is an emergent property arising from many-body effects in the complex assembly, rather than from any specific pairwise interaction. From a theoretical perspective, this behavior can be rationalized by the concept of spinodal decomposition[56, 57]. As illustrated schematically in **Figure 2C**, we consider the free energy as a function of the particle number of a single subunit embedded in the full assembly. For a convergent subunit, its experimental abundance lies within a convex region of the free-energy well, where the corresponding chemical potential—given by the local slope—is well-defined and stable. By contrast, for a divergent subunit, the discontinuity and non-invertibility of the abundance–chemical potential relation indicate that its empirical abundance resides within a non-convex region of the free-energy landscape associated with the transition between dilute and dense phases. In this spinodal-like regime, the grand-canonical ensemble becomes ill-posed, and inference of a unique chemical potential necessarily fails. The observed divergence is therefore a physical signature of thermodynamic instability, not an artifact of the inference procedure.

These convergent and divergent behaviors are not strictly binary but lie on a continuum. In our analysis, Arp5 at a volume fraction of 0.1 and Nhp10 at a volume fraction of 0.2 exhibit critical behavior at the boundary between convergent and divergent regimes. The microscopic origins of these critical phenomena remain to be elucidated and will require further investigation. Nonetheless, by combining realistic structural and abundance data with a grand-canonical description of a crowded, frustrated assembly, our study advances the understanding of general physical principles governing spatial organization of megacomplexes in relation to cell state.

### B. Divergent subunits lead to assembly of clusters whose subunit composition and spatial arrangement vary with crowding

Here, we focus on the INO80 chromatin-remodeling complex, which is characterized by a dynamic and heterogeneous composition that cannot be captured by a single static structure. INO80 exists in multiple compositional states whose subunit diversity is tightly linked to its functions, including its well-established roles in regulating transcription and cellular metabolism. Despite the importance of this complex, the mechanistic contributions of several key subunits, such as Nhp10, Ies4, and Ies6, have remained unresolved at a structural and thermodynamic level. In our framework, we identify these subunits as divergent: their behavior changes qualitatively with crowding, and their emergent divergence across volume fractions leads to distinct cluster assemblies that differ in subunit composition and spatial organization. **Figure 4B** summarizes the divergent subunits and the clusters they form at each volume fraction (see also **Figure S6**). The resulting visualizations highlight clear morphological differences in assembly architecture across crowding conditions, shaped jointly by protein–protein interactions and abundances. We speculate that these emergent clusters form the basis for remodeling INO80 into large megacomplexes that engage DNA and actin filaments in multiple configurations.

Notably, every cluster we observe contains at least one divergent subunit, suggesting that divergence plays a critical role in cluster nucleation and stabilization. Consistent with this, the potentials of mean force (PMFs) in **Figure 2B** show that divergent subunits act as effective "stickers," and **Figure 3** demonstrates that they strongly modulate global thermodynamic parameters of the assembly, whereas convergent subunits have a comparatively minor impact. We therefore propose that divergent subunits promote nucleation through long-range, many-body interactions that emerge from the collective distribution of subunit interaction strengths and abundances. In this sense, divergent subunits function as "coordinators" of spatial organization, while convergent subunits serve as "participants" whose behavior is largely shaped by the surrounding environment. Although our model does not yet resolve atomic-level details of INO80 assembly, it provides a mechanistic framework for understanding how subunit-specific characteristics and macromolecular crowding jointly drive the remodeling dynamics and megacomplex formation of INO80.

### C. A machinery view of INO80 remodeling DNAs with divergent subunits that take cues from crowding

With a mechanistic view of how crowding shapes molecular assemblies by invoking the divergent subunits, we propose a mechanism for the INO80, a Chromatin Remodeling Complex (CRC), to anneal the shape and position of the DNA fragment it binds to (**Figure 4C**). We liken INO80 to a toolbox, in which each module acts as a "tool" that executes a task. At low volume fractions $\phi = 0.05$, these tools may be loosely connected with each other. We labeled them with transparent colors. As the volume fraction increases to $\phi = 0.1$, some subunits, for example Arp5 (purple) and Ies6 (purple), become divergent. We labeled them with opaque colors. The Arp5 and Ies6 subunits promote the incorporation of **Arp5** module into the assembled machinery into the INO80 CRC complex. In contrast, the state of **Nhp10** module and **Arp8** module remains unchanged as crowding increases from $\phi = 0.05$ to 0.1. As $\phi$ increases from 0.1 to 0.2, the Nph10 subunit (blue) becomes divergent, and the Nph10 module is further recruited to the machinery of INO80 CRC. Interestingly, although only Ies6 remains divergent, the **Arp5** module still remains in the assembled CRC, since it only takes one divergent subunit to assemble a module.

These observations manifested a set of programmable units to manipulate INO80 CRC, whose biological function is to remodel the position of a DNA fragment powered by its ATPase domain. We demonstrated that the inner workings of INO80 subunits are influenced by crowding. Among the known subunits of INO80, the structure of Nhp10 module is intrinsically ordered and unknown. We speculate that its position as a peripheral module could act as a flexible binding target to be recruited by other CRC complexes to form even larger megacomplexes through a "fly-casting mechanism[58-60]".

This perspective considers crowding as an active regulatory mechanism for organizing higher-order complex formation. The intracellular environment is inherently crowded [61, 62], and macromolecular concentrations significantly influence collective behavior through excluded-volume effects [48, 62]. Cells are believed to maintain crowding homeostasis to ensure proper function [38, 39, 63]. Previous computational studies have examined the regulatory role of crowding by quantifying its contribution to the system's free energy [64]. Our approach expands on this view by treating volume fraction and accessible volume as thermodynamic conjugate variables [53], similar to the relationship between pressure and volume in mechanical work, with volume fraction effectively proportional to hydrostatic pressure. Unlike active processes that require ATP and are limited by energy supply, crowding can spontaneously influence megacomplex formation in response to physical parameters such as osmotic pressure, cell volume, and macromolecular concentration [38, 39]. In this framework, divergent subunits may act as molecular points of crosstalk that sense these environmental cues and initiate higher-order spatial organization of INO80 at the appropriate time and place.

**Acknowledgments**
JW, JN, and JH thank the support from the National Science Foundation MCB 2221824. This work is also partially supported by the NW-BRaVE for Biopreparedness project funded by the U. S. Department of Energy (DOE), Office of Science, Office of Biological and Environmental Research, under FWP 81832. A portion of this research was performed on a project award (Enhancing biopreparedness through a model system to understand the molecular mechanisms that lead to pathogenesis and disease transmission ) from the Environmental Molecular Sciences Laboratory, a DOE Office of Science User Facility sponsored by the Biological and Environmental Research program under Contract No. DE-